\def\mearth{{\rm\,M_\oplus}}

\def\gsim{~\rlap{$>$}{\lower 1.0ex\hbox{$\sim$}}}
\def\lsim{~\rlap{$<$}{\lower 1.0ex\hbox{$\sim$}}}

\def\h2o{\rm{H_{2}O}}
\def\mh2{\rm{H_{2}}}

\def\co2{\rm{CO_{2}}}
\def\ch4{\rm{CH_{4}}}
\documentclass[12pt, preprint]{aastex}
\usepackage{graphicx}
\usepackage{natbib,color,lscape}
\usepackage{subfigure}
\usepackage{threeparttable}
\usepackage{url}
\usepackage{subfigure}
\usepackage{threeparttable}
\usepackage{rotating}

\begin{document}
\title{A REVISED ESTIMATE OF THE OCCURRENCE RATE OF TERRESTRIAL PLANETS IN THE HABITABLE ZONES
 AROUND {\it Kepler} M-DWARFS  }
\author{Ravi kumar Kopparapu\altaffilmark{1,2,3,4}} 
\altaffiltext{1}{Department of Geosciences, Penn State University, 443 
Deike Building, University Park, PA 16802, USA}
\altaffiltext{2}{NASA Astrobiology Institute's Virtual Planetary Laboratory}
\altaffiltext{3}{Penn State Astrobiology Research Center, 2217 Earth and Engineering Sciences Building
University Park, PA 16802}
\altaffiltext{4}{Center for Exoplanets \& Habitable Worlds, The Pennsylvania State University, University
Park, PA 16802}

\begin{abstract}

Because of their large numbers, 
 low mass stars may be the most abundant planet hosts in our Galaxy. Furthermore, terrestrial planets
in the habitable zones (HZs) around  M-dwarfs can potentially be characterized in the
near future
and hence may be the first such planets to be studied.
Recently  \cite{DC2013} used {\it Kepler} data and
calculated the  frequency of terrestrial planets in the HZ of  cool stars to be
 $0.15^{+0.13}_{-0.06}$ per star for Earth-size
planets ($0.5-1.4$ R$_{\oplus}$). However, this estimate was derived using the \cite{Kasting1993} HZ 
limits, which were not valid for stars with effective temperatures lower than $3700$ K. 
Here we update their result using new HZ limits from \cite{Kopp2013} for stars with effective temperatures
between $2600$ K and $7200$ K, which includes the cool M stars in the {\it Kepler} target list. 
The new habitable zone boundaries  increase the number of planet 
candidates in the habitable zone.
Assuming Earth-size planets as $0.5 - 1.4$ R$_{\oplus}$, when we reanalyze their results, 
 we obtain a terrestrial planet frequency of $0.48^{+0.12}_{-0.24}$ and
$0.53^{+0.08}_{-0.17}$ planets per M-dwarf star
for conservative and optimistic limits of the HZ boundaries, respectively.
Assuming Earth-size planets as
$0.5 - 2$ R$_{\oplus}$, the  frequency increases to
$0.51^{+0.10}_{-0.20}$ per star for the conservative estimate and to $0.61^{+0.07}_{-0.15}$ per star
for the optimistic estimate.  
Within uncertainties, our optimistic estimates 
are in agreement with a similar optimistic
estimate from the radial velocity survey of M-dwarfs ($0.41^{+0.54}_{-0.13}$, \cite{Bonfils2011}).
 So, the potential for 
finding Earth-like planets around M stars may be higher than previously reported.

\end{abstract}

\maketitle

\section{Introduction}
\label{intro}
Advances in detection techniques and recent exoplanet surveys have 
discovered  terrestrial planets in the habitable zone (HZ) of their parent star
\citep{Udry2007, Vogt2010, Pepe2011a, Borucki2011, Bonfils2011, Borucki2012, Vogt2012, Tuomi2012b,
AE2012} and it is
expected that this number will greatly increase as time passes \citep{Batalha2012}.
Classically, the HZ is defined as the circumstellar region in which a
terrestrial mass planet ($0.1 \lesssim M \lesssim 10 \mearth$), with favorable
atmospheric conditions, can sustain liquid water on its surface
\citep{Huang1959, Hart1978, Kasting1993, Underwood2003, Selsis2007b, KS2011, Kopp2013}. 
Currently, more than 800 extra-solar planetary systems have been
detected\footnote{\url {exoplanets.org}}, and  $\sim 3000$ additional candidate systems
from the {\it Kepler} mission are waiting to be confirmed \citep{Batalha2012}.

One of the primary goals of {\it Kepler} mission is to determine $\eta_{\oplus}$,
the frequency of Earth-size planets in and near the HZ of solar-type stars \citep{Borucki2011}.
Such estimates have been made for potentially rocky planets using both radial-velocity 
(RV,  \cite{Bonfils2011}) 
%(RV, \citep{Howard2010, Bonfils2011, Mayor2011}, Swift2012) 
and {\it Kepler} data \citep{CS2011, Traub2012, DC2013}.
%and {\it Kepler} data \citep{CS2011,Howard2012, Traub2012, DC2013}.
Furthermore, many other studies have estimated in general the terrestrial planet frequency 
\citep{Howard2010, Mayor2011, Howard2012, Swift2012} and 
the consensus from these studies is that there are more low mass/smaller radii planets than high mass/larger
radii ones. 
Moreover, the frequency increases towards lower radii, larger orbital periods \citep{Howard2012}.
The estimates of the occurrence rates of terrestrial planets  with
periods $< 50$ days ranged from $0.23^{+0.16}_{-0.1}$ \citep{Howard2010} around GKM stars using RV,  to 
$0.51^{+0.06}_{-0.05}$ \citep{DC2013} around M-stars using transit detection. 
%\cite{Bonfils2011} obtained a value of $0.41$ However, \cite{Traub2012} 

Specifically, to estimate $\eta_{\oplus}$, 
one needs to know the
boundaries of the HZs. Some studies  did estimate this number for M-dwarfs
using transit ($0.15^{+0.13}_{-0.06}$, \cite{DC2013}) and RV ($0.41^{+0.54}_{-0.13}$, \cite{Bonfils2011}) 
surveys.
\cite{Traub2012} estimated  $\eta_{\oplus}$ to be $0.34 \pm 0.14$ for {\it Kepler} FGK stars, but this is 
based on 
an extrapolation of data for planets with orbital periods shorter than 42 days.
These estimates of the occurrence rates were based on
 1-D radiative-convective, cloud-free climate model 
calculations by \cite{Kasting1993}. Several other studies \citep{Underwood2003, Selsis2007b} 
parametrized \cite{Kasting1993} results to estimate relationships between HZ boundaries and stellar parameters 
for stars of different spectral types. Moreover, no uniform criterion of HZ limits were followed in applying them
to calculate the occurrence rates,  which can lead to comparing quantities that are not similar. 
For example, for the outer edge of the HZ, some studies \citep{DC2013} used the first $\co2$ condensation limit 
and others \citep{Bonfils2011} used Early Mars criterion.

Recently \cite{Kopp2013} obtained new, improved estimates of the boundaries of the HZ by 
updating \cite{Kasting1993} model with new $\h2o$ and $\co2$ absorption coefficients 
from updated line-
by-line (LBL) databases such as HITRAN 2008  \citep{Rothman2009} and HITEMP 2010 \citep{Rothman2010}.
They estimated that, for our Sun,  a conservative estimate of the inner HZ (IHZ) is at
$0.99$ AU and the outer HZ (OHZ) is at $1.70$ AU. These values represent the 
``water loss'' (moist greenhouse) limit at the inner edge and ``maximum greenhouse'' limit at the outer edge.

\cite{Kopp2013} have also estimated HZ boundaries around other stellar spectral types, including  M stars,
 which are primary targets for ongoing
surveys such as {\it Habitable Zone Planet Finder} \citep[HPF]{Suvrath2012} and MEarth 
 \citep{Nutzman2008} to discover potential habitable planets. Furthermore, M-dwarfs are also increasingly
becoming important as {\it Kepler} targets to find terrestrial size planets: 
The planetary orbital periods around these stars are shorter, 
the transit signal is larger, and there is a greater chance of detecting these planets than there is around
 a Sun-like star.
Although M-dwarfs are fainter, the HZs are also closer for M-dwarfs, so it is likely that 
potential habitable planets may be discovered around nearby 
M-dwarfs in the near future with space-based characterization
missions.

In this study, we apply \cite{Kopp2013} HZ limits to estimate the occurrence rate of terrestrial size 
($0.5 - 1.4$ R$_{\oplus}$ and  $0.5 - 2$ R$_{\oplus}$) planets in the HZ of M stars. We base our analysis on \cite{DC2013} who updated
stellar parameters of 3897 low mass  {\it Kepler} target stars with temperatures below $4000$ K using
Darthmouth stellar evolutionary models \citep{Dotter2008, Feiden2011}. 
	 The outline of the paper is as follows: In \S\ref{sec2} we will revise \cite{DC2013} calculations
with new HZ estimates of \cite{Kopp2013}, discuss the implications in \S\ref{sec3} and conclude in 
\S\ref{conclusions}.

\section{Estimate of habitable zone planet occurrence rate around M-dwarfs}
\label{sec2}

From \cite{Howard2012} and \cite{DC2013},
the planet occurrence rate over a given period ($P$)
and radius ($R_{p}$) range is given by:
 
\begin{eqnarray}
f(R_{p},P) &=& \sum_{i=1}^{N_{p}(R_{p},P)} \frac{a_{i}}{R_{\star,i} N_{\star,i}}
\label{rateeq}
\end{eqnarray}
where $a_{i}$ is the semi-major axis of planet $i$, $R_{\star,i}$ is the host star's radius of planet
$i$, $N_{\star,i}$ is the number of stars around which planet $i$ could have been detected and 
$N_{p}(R_{p},P)$ is the number of planets with the radius $R_{p}$ and period $P$.
The ratio $a_{i}/R_{\star,i}$ is the inverse of the probability of transit orientation, which is considered to
take non-transiting geometries into the estimation of occurrence rate. 

In Table 2 of \cite{DC2013}, the authors provide stellar and planetary parameters of candidate KOIs that are
considered to be in or near the HZ. To estimate the occurrence rate, \cite{DC2013} adopt the most
conservative estimate of HZ limits from \cite{Kasting1993} results: the `moist greenhouse' for the
inner HZ (0.95 AU) and the beginning of the $\co2$ condensation for the outer HZ (1.37 AU for the Sun). 
Furthermore, they consider 'Earth-size' as planets that are in the radius range $0.5 - 1.4$ R$_{\oplus}$.
 Based on these definitions,  \cite{DC2013} consider two KOIs to be in the HZ from their Table 2:
KOI  2626.01 and 1422.02.  They then calculate occurrence rate of Earth-size planets in the 
HZ as $0.15^{+0.13}_{-0.06}$ planets per star.

This estimate may need to be updated, however, as the estimated width and the position 
of the HZ has changed recently, following new calculation by \cite{Kopp2013}\footnote{\url {http://www3.geosc.psu.edu/~ruk15/planets/}}.
%However, recently \cite{Kopp2013} derived new HZ limits\footnote{\url {http://www3.geosc.psu.edu/~ruk15/planets/}}
% around main-sequence stars. 
According to these
authors, HZs are  farther out from their star than what has been calculated previously. This will
have a significant effect on the occurrence rate derived by \cite{DC2013}. 
Also, The ``1st $\co2$ condensation'' limit of \cite{Kasting1993}, should now be disregarded, as it
has been shown that $\co2$ clouds generally warm a planet's climate \citep{FP1997}).
Here we use the 
\cite{Kopp2013} HZ limits to derive revised estimates of occurrence rates of potentially habitable 
planets around low-mass stars.  In order to put a lower and upper bound on $\eta_{\oplus}$, we calculate
 two values for the occurrence rate based on the conservative and optimistic estimate of HZ limits as given
in \cite{Kopp2013}:

(1) In a conservative estimate, the inner edge of the HZ is determined by the ``moist-greenhouse''
limit which is derived by assuming a fully saturated troposphere and negligible cloud feedback.
Neither assumption is likely true in reality, but it is  difficult to improve on this with
 a 1-D climate model because such models cannot accurately simulate clouds or relative humidity.
 The outer edge of the HZ 
is determined by the ``maximum greenhouse'' limit where a $\co2$ dominated atmosphere can produce
maximum amount of greenhouse warming. Here also, the radiative warming by $\co2$ clouds is neglected
hence the limit is a conservative estimate.

(2) In an  optimistic scenario, the inner edge of the HZ boundary can be obtained by 
the ``recent Venus'' limit which is based on the observations of Venus by Magellan spacecraft,
 suggesting that liquid water has been absent from the surface of Venus for at least
1 Gyr \citep{SH1991} or earlier. 
The Sun at that time was $\sim92 \%$ of the present day
luminosity, according to standard stellar evolutionary models \citep[See Table 2]{Baraffe1998, Bahcall2001}.
The current solar flux at Venus distance is $1.92$ times that of Earth.
 Therefore, the solar flux received by Venus at that time was $0.92 \times 1.92 = 1.76$ times that of Earth.
This empirical estimate of the inner HZ edge in our Solar System corresponds to an
orbital distance of $d=(1/1.76)^{0.5} = 0.75$ AU for the present day. Note that this distance
is greater than Venus' orbital distance of 0.72 AU because the constraint of surface water 
was imposed when the Sun was fainter.
%at an earlier time in the planet's history.
The  outer edge optimistic estimate is the ``Early Mars'' limit based on the
observation that
early Mars was warm enough for liquid water to flow on its surface \citep{Pollack1987, Bibring2006}.
Assuming the dried up riverbeds and valley networks on martian surface are 3.8 Gyr old, the solar luminosity at
that time would have been $\sim 75 \%$ of the present value (See Eq.(1) in \cite{Gough1981} and Table 2
in \cite{Bahcall2001}). The present-day solar flux at Mars distance is $0.43$ times that of Earth. Therefore,
 the solar flux received by Mars at 3.8 Gyr was $0.75 \times 0.43 = 0.32$ times that of Earth.
 The corresponding empirical OHZ limit for our Solar System today, then, would be
$d = (1/0.32)^{0.5} \approx 1.77$ AU.

The optimistic HZ limits 
%(recent Venus at the inner edge and Early Mars at the outer edge) 
can be 
extrapolated to other stellar types with effective temperatures between
$2600$ K to $ 7200$ K, by scaling them with the 
corresponding values of conservative HZ limits,
%(moist-greenhouse at the inner edge and maximum greenhouse
%at the outer edge), 
as shown in \cite{Kopp2013}.
%Both the conservative and optimistic estimates are given in 
%a parametric form by \cite{Kopp2013} that can be applicable to stars with effective temperatures between
%$2600$ K to $ 7200$ K. 

To derive new rates from Eq.(\ref{rateeq}), we need two quantities: (1) $a_{i}/R_{\star,i}$ for the 
planets in the HZs and (2) $N_{\star,i}$, the number of stars around which a planet that
has the same size and receives the same insolation as those KOIs in the HZ
could have been
detected.  The photometrically derived
values of $a_{i}/R_{\star,i}$  are given in Table 5 of \cite{DC2013}. But 
\cite{DC2013} use the calculated $a_{i}/R_{\star,i}$ determined from the period of the planet 
and mass of the star (Dressing, private communication). 
% As for $N_{\star,i}$, we do not
%have  sufficient information to calculate the number of stars with planets that receive the same 
%insolation. Instead, 
We use the $N_{\star,i}$ values provided within the period-radii cells of Fig. 15 of 
\cite{DC2013}, which gives the number of stars around which a planet from the
center of the grid cell would have been detected with a signal to noise ratio above 7.1 $\sigma$. 
 This should still give us nearly the same occurrence rate, or  an underestimate of
 the actual value (see next section). 
%But we think it will not be vastly different from the actual rate estimate if appropriate numbers are used. 
For example, for the two KOIs (2626.01 and 1422.02) that \cite{DC2013} consider to be in the HZ, when we 
use the center of the grid cell $N_{\star,i}$ numbers (1822 and 872) from Fig. 15 of \cite{DC2013} (instead of
the $N_{\star,i}$ values provided in section 5.7) 
and use the period determined $a_{i}/R_{\star,i}$ as they did, we were able to reproduce their value of $0.15$. 
Thus, our new occurrence rate (which uses center of the grid cell numbers) determined here
probably is close to a value from a more rigorous estimate.

We will first calculate $\eta_{\oplus}$  using \cite{DC2013}
radius range of $0.5 - 1.4$ R$_{\oplus}$. 
Assuming conservative HZ limits\footnote{ie., Moist-greenhouse limit at the inner edge and maximum 
greenhouse limit at the outer edge} from \cite{Kopp2013}, four KOIs from Table 2 of \cite{DC2013} should be
 in the HZ based on the insolation fluxes (Fig.\ref{koihz}):
KOI 1686.01, 2418.01, 2626.01 and 1422.02. 
We use the photometric $a_{i}/R_{\star,i}$ from Table 5
of \cite{DC2013}, rather than the calculated  $a_{i}/R_{\star,i}$ from the period and stellar mass. 
For a typical planet  candidate, the photometric 
$a_{i}/R_{\star,i}$  is 85$\%$ of the $a_{i}/R_{\star,i}$ from the period and stellar
mass. Therefore, our value tends to be at the lower end of the occurrence rate estimate.
The corresponding $a_{i}/R_{\star,i}$ for these
candidates from Table 5 of \cite{DC2013} are: $102.482$, $116.837$, $36.283$ and $51.985$, 
respectively. The corresponding $N_{\star,i}$ values from center of the grid cells of Fig.15 
are: 353, 994, 1822 and 872, respectively.
Using Eq.(\ref{rateeq}), we get the conservative estimate of $\eta_{\oplus}$
for low-mass stars to be $0.48^{+0.12}_{-0.24}$ per star. 

An optimistic estimate of the occurrence rate can also be derived based on the recent Venus and Early 
Mars limit from \cite{Kopp2013} results. 
Assuming these limits, six KOIs from Table 2 of \cite{DC2013} should be in the HZ (Fig.\ref{koihz}):
KOI  1686.01, 2418.01, 2626.01, 1422.02, 2650.01 and 886.03. 
 The corresponding photometric determined $a_{i}/R_{\star,i}$ for these candidates from Table 5 of
\cite{DC2013} are:
$102.482$, $116.837$, $36.283$, $51.985, 54.052$ and $39.246$.  The $N_{\star,i}$ values from  Fig.15 of
 \cite{DC2013} are: 353, 994, 1822,  872, 1822 and 2336. 
Using Eq.(\ref{rateeq}), the optimistic estimate of the
occurrence rate of Earth-size planets in the habitable zones around low-mass stars is $0.53^{+0.08}_{-0.17}$ 
per star.

Instead of assuming $0.5 - 1.4$ R$_{\oplus}$ as `Earth-size', we will also calculate occurrence rate 
extending the radius range from $0.5 - 2$ R$_{\oplus}$. Planets with radius $> 1.4$ R$_{\oplus}$ 
are thought to have either homogeneous composition of water ice, silicate or iron or some
differential composition of these compounds \citep{Seager2007, Rogers2011, Lopez2012}.
 This will then add KOI $854.01$ in the 
conservative rate estimate (the total number of KOIs in the HZ is then five) and KOI $250.04$ 
in the optimistic rate estimate (the total HZ KOIs is eight).
The corresponding $a_{i}/R_{\star,i}$ and $N_{\star,i}$ for these additional
candidates from \cite{DC2013} Table 5  and Fig. 15 are: $90.045$ and $2991$, respectively, for KOI $854.01$;
 $157.259$ and $3287$, respectively, for KOI $250.04$.

Using this expanded definition of Earth-size ($0.5 - 2$ R$_{\oplus}$), a conservative estimate of
the occurrence rate of Earth-size planets in the HZs around M-dwarfs is $0.51^{+0.10}_{-0.20}$ per star.
An optimistic estimate on the occurrence rate is $0.61^{+0.07}_{-0.15}$ per star. 
%Note that the uncertainties are 
%overestimated, as discussed above, and should be much less than our calculated values.

\section{Discussion}
\label{sec3}
\begin{figure}[!hbp|t]
\includegraphics[width=0.95\textwidth]{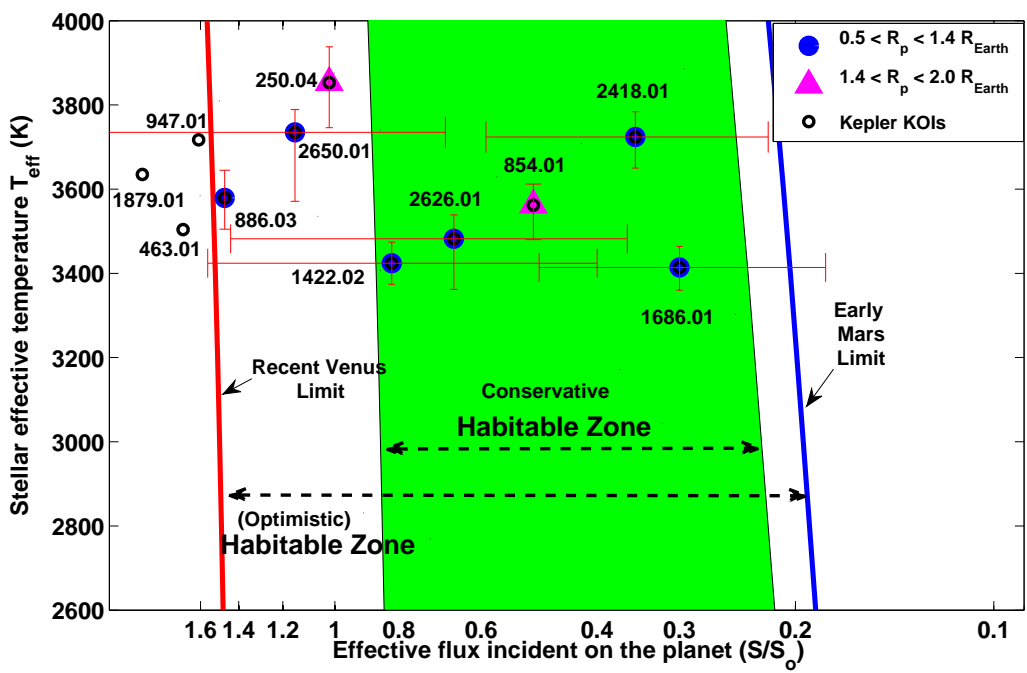}
\caption{Incident stellar flux on a planet as a function of stellar effective temperature, 
$T_{eff}$. The green shaded region is the conservative HZ. The optimistic HZ limits are recent
Venus (solid red curve) and Early Mars (solid blue curve). Two of the 
terrestrial-size KOIs (1422.02 and 2626.01) that are in the  \cite{DC2013} HZ are also shown.} 
\label{koihz}
\end{figure}

The occurrence rate estimates derived in the previous section indicate that terrestrial size planets 
 in HZs around low-mass stars are more frequent than previously thought. 
Couple of caveats are to be noted: 

(1) The KOIs  are not confirmed planets. The
calculated false positive rates for {\it Kepler} candidates in our relevant radius bin range from
$12.3\%$ \citep{Fressin2013} to $\sim 10\%$ \citep{MJ2011}.  Therefore, 
we may be overestimating $\eta_{\oplus}$. Furthermore, the uncertainties on the fluxes are large
(Fig. \ref{koihz}), so some may not be in the HZ (2) On the other hand, We use photometrically 
derived $a_{i}/R_{\star,i}$ given in Table 5 of \cite{DC2013}. This may {\it underestimate} our
occurrence rate because \cite{DC2013} use calculated $a_{i}/R_{\star,i}$ from orbital periods and stellar mass
(Dressing 2013, private communication) and the photometric values are $\sim 85\%$ of the derived values.
So the net effect from points 1 and 2 may not change significantly our estimate of $\eta_{\oplus}$.

 (3) As mentioned above, we do not calculate number of stars 
around which a planet that has the same size and receives the same insolation as the ones in the HZ could have 
been detected; we instead use
the cell numbers from Fig.15 of \cite{DC2013}, which gives the number of stars around which a 
hypothetical planet that has the same radius and period as the
center of the grid cell would have been detected with a signal to noise ratio above $7.1 \sigma$.  There is 
no reason to assume that the KOIs considered here to be in the HZ are at the center of the grid cell (in fact,
they are not). This
offset  probably overestimates our value of $N_{\star,i}$ in Eq.(\ref{rateeq}).
% because
%most of those planet candidates orbit stars that are cooler than the typical Kepler M dwarf.

 For example, in
section 5.7 of \cite{DC2013}, they
calculate $N_{\star,i}$ values of 2853, 813 and 2131 for KOIs $854.01, 1422.02$ and $2626.01$, respectively. 
These numbers are 
generally lower in value than the center of the grid cells values we use from Fig.15 
($2991, 872, 1822$, respectively), except for KOI 2626.01 (an increase of $~17\%$).
If this is the general trend, i.e, if we are
overestimating $N_{\star,i}$ systematically, then our calculated occurrence rates of terrestrial planets in the
HZ around M-dwarfs can be considered as a {\it lower} bound to the actual value. 
%But, as pointed in (2), if we are underestimating both
%$a_{i}/R_{\star,i}$ (because we are using photometric values from Table 5 of \cite{DC2013})
%and $N_{\star,i}$, then the occurrence rate (from Eq.(\ref{rateeq})) would probably be 
%close to our estimated value, within uncertainties. 
Nevertheless, we tested to see how much the rate would change if we change $N_{\star,i}$.
It should be noted that $N_{\star,i}$ (at same insolation) is most likely smaller than 
$N_{\star,i}$ (at same period) for the planet
candidates in the HZ because most of those planet candidates orbit stars that are cooler than the
typical Kepler M dwarf.
Since there is one KOI (2626.01) for which we underestimate $N_{\star,i}$ by $\sim 17\%$, and could
potentially decrease our occurrence rate estimate, we rounded off and added  $20\%$ of the respective
$N_{\star,i}$ values to {\it all} the KOIs that are in the  HZ. After performing this calculation, 
for the radius range $0.5 - 1.4$ R$_{\oplus}$, the conservative estimate changed from $0.48^{+0.12}_{-0.24}$ to 
$0.41^{+0.10}_{-0.20}$ per star. The optimistic estimate  changed from $0.53^{+0.08}_{-0.17}$
to $0.44^{+0.07}_{-0.14}$ per star. 
In the extended radius range for Earth-size ($0.5 - 2.0$ R$_{\oplus}$), the conservative estimate 
changed from $0.51^{+0.10}_{-0.20}$ per star to $0.43^{+0.08}_{-0.17}$, and the optimistic estimate
changed from $0.61^{+0.07}_{-0.15}$ per star to $0.52^{+0.06}_{-0.13}$.
%As explained before, the uncertainties are overestimated.

Note that even though adding $20\%$  to $N_{\star,i}$ lowers our occurrence rate, we are underestimating
$a_{i}/R_{\star,i}$ on an average of $~15\%$ or more.
Furthermore, we showed that we were able to reproduce 
\cite{DC2013} estimate even if we use center of the grid cell value for $N_{\star,i}$. Therefore, a more
rigorous analysis to determine the occurrence rate from \cite{DC2013} would likely produce a similar
result as our's.

\cite{Bonfils2011} studied $102$ southern nearby M dwarfs using ESO/HARPS spectrograph and obtained the 
frequency of terrestrial mass planets ($1 - 10$ M$_{\oplus}$) in the HZ to be $0.41^{+0.54}_{-0.13}$.
Note that they consider two planets to be in the HZ, Gl 581d and Gl 667Cc, using optimistic
limits of the HZ (recent Venus and Early Mars) from \cite{Selsis2007b}. 
If we use also use  optimistic HZ estimates, we find that both the planets are in the  HZ. 
Our optimistic estimate of $\eta_{\oplus}$ ranges from $0.53^{+0.08}_{-0.17}$ to $0.61^{+0.07}_{-0.15}$ per star,
depending on if the mass range in \cite{Bonfils2011} is applicable to terrestrial sizes of 
$0.5 - 1.4$ R$_{\oplus}$ radius or $0.5 - 2$ R$_{\oplus}$ radius. Nevertheless, our estimates of 
the occurrence rates of HZ terrestrial planets around M-dwarfs are in
good agreement with independently derived RV estimates, within the error bars.

\thispagestyle{empty}

%The \cite{Bonfils2011} estimate of 
%frequency of terrestrial planets in the HZ then drops
%down to $0.20$ from $0.41$.
% This is lower in value compared to our conservative estimates, most likely because of
%small number of stars that are being sampled. 
%The fact that the estimated $\eta_{\oplus}$ value changes by a 
%factor of two when a single planet moves out of the HZ indicates that the statistics on which this estimate is 
%based are far from providing a definitive answer. 

	\section{Conclusions}
	\label{conclusions}
The purpose of our analysis is to 
outline the significant increase in the terrestrial planet occurrence rate in the HZs of M-dwarfs compared
to \cite{DC2013}. 
	We have obtained revised estimates based on \cite{DC2013} estimates of 3987 
{\it Kepler} M-dwarfs that are cooler than
$4000$ K in Q1-Q6 data.  Applying new HZ results from \cite{Kopp2013} to Earth-size planets 
($0.5 - 1.4$ R$_{\oplus}$),  we calculate that a conservative estimate of the occurrence
rate of Earth-size planets in the HZs around M-dwarfs is $0.48^{+0.12}_{-0.24}$ planets per star.  
The optimistic estimates indicate that the occurrence rate increases to $0.53^{+0.08}_{-0.17}$ planets per star. 
If we extend the definition of Earth-size to planets in the radius range $0.5 - 2.0$ R$_{\oplus}$, the 
conservative estimate increases to $0.51^{+0.10}_{-0.20}$ per star, and the optimistic estimate increases
to $0.61^{+0.07}_{-0.15}$ per star.
As discussed in our paper, our 
results probably are close or underestimate the actual occurrence rate of Earth-size planets in the HZs around
M-dwarfs. 
%The uncertainties in our values are overestimated, and the actual error bars should be lower because
%of inclusion of additional HZ planet candidates in our occurrence rate estimate. 
Furthermore, our optimistic
value of $\eta_{\oplus}$ quoted above is in agreement (within uncertainties) with a similar estimate,
 $0.41^{+0.54}_{-0.13}$, from \cite{Bonfils2011} ESO/HARPS survey of 102 M-dwarf stars indicating that
the frequency of terrestrial planets in the HZs of M-dwarfs may be higher than previously reported.

	\acknowledgements

        The author is grateful to Courtney Dressing for discussions leading to this paper.
	The author would also like to thank James Kasting, Steinn Sigurdsson, 
Eric Feigelson, Suvrath Mahadevan,
Jason Wright, 
        Chester Harman, Ramses Ramirez for their valuable input and an 
anonymous referee whose comments 
 improved the manuscript. 
	R. K,  gratefully acknowledge funding from NASA Astrobiology
	 Institute's  Virtual 
	Planetary Laboratory lead team, supported by NASA under cooperative agreement
	NNH05ZDA001C, and the Penn State Astrobiology Research Center.

\end{document}